\documentclass[%
 reprint,showkeys,
 amsmath,amssymb,
 aps,
]{revtex4-2}

\usepackage{graphicx}% Include figure files
\usepackage{dcolumn}% Align table columns on decimal point
\usepackage[usenames]{color}
\usepackage{colortbl}
\usepackage{bm}% bold math
\usepackage{longtable}
\begin{document}

%\preprint{APS/123-QED}

\title{Photonuclear reactions ${^{65}\rm{Cu}}(\gamma,n)^{64}\rm{Cu}$
	and ${^{63}\rm{Cu}}(\gamma,xn)^{63-x}\rm{Cu}$ cross-sections \\
	in the energy range \mbox{$E_{\rm{\gamma max}}$ = 35--94 MeV} } 

\author{O.S. Deiev, I.S. Timchenko}
%\author{}%
 \email{timchenkooryna@gmail.com}
%\affiliation{%  Authors' institution and/or address\\  This line break forced with \textbackslash\textbackslash }%


\author{S.M. Olejnik, S.M. Potin, \\
	V.A. Kushnir, V.V. Mytrochenko, S.A. Perezhogin, B.I. Shramenko}
\affiliation{ National Science Center "Kharkov Institute of Physics and Technology", \\
 1, Akademichna St., 61108, Kharkiv, Ukraine}%

\date{\today}% It is always \today, today,

\begin{abstract}
The bremsstrahlung flux-averaged cross-sections $\langle{\sigma(E_{\rm{\gamma max}})}\rangle$ for the photonuclear reactions ${^{65}\rm{Cu}}(\gamma,n)^{64}\rm{Cu}$, ${^{63}\rm{Cu}}(\gamma,n)^{62}\rm{Cu}$, ${^{63}\rm{Cu}}(\gamma,2n)^{61}\rm{Cu}$ and ${^{63}\rm{Cu}}(\gamma,3n)^{60}\rm{Cu}$ have been measured in the range of bremsstrahlung end-point energies $E_{\rm{\gamma max}}$ = 35--94 MeV.
	 The experiments were performed with the electron beam from the NSC KIPT linear accelerator LUE-40 with the use of the activation and off-line $\gamma$-ray spectrometric technique. The calculation of the flux-average cross-sections $\langle{\sigma(E_{\rm{\gamma max}})}\rangle_{\rm{th}}$ was carried out using the cross-section $\sigma(E)$ values from the TALYS1.95 code with the default options.\\
	 It is shown that the experimental average cross-sections for the reactions ${^{65}\rm{Cu}}(\gamma,n)^{64}\rm{Cu}$, ${^{63}\rm{Cu}}(\gamma,n)^{62}\rm{Cu}$, ${^{63}\rm{Cu}}(\gamma,2n)^{61}\rm{Cu}$ are systematically higher than the theoretical estimates in the TALYS1.95 code. The obtained $\langle{\sigma(E_{\rm{\gamma max}})}\rangle$ supplement the data of different laboratories for the case of reactions $(\gamma,n)$ and $(\gamma,2n)$. For the reaction ${^{63}\rm{Cu}}(\gamma,3n)^{60}\rm{Cu}$, the values of $\langle{\sigma(E_{\rm{\gamma max}})}\rangle$ were measured for the first time.
\end{abstract}


%Use showkeys class option if keyword
\maketitle

%\tableofcontents

\section{Introduction}
\label{intro}

Most of the data on cross-sections for photonuclear reactions, which are important for many fields of science and technology, as well as for various data files (EXFOR, RIPL, ENDF, etc.), were obtained in experiments using quasi-monoenergetic annihilation photons in the energy range of the giant dipole resonance (GDR). As a rule, such experiments were carried out with the direct detection of neutrons from the output channel of the reaction. However, when measuring the neutron yield, it becomes difficult to separate neutrons registered from various reactions $(\gamma,n)$, $(\gamma,pn)$, $(\gamma,2n)$, $(\gamma,p2n)$, $(\gamma,3n)$, etc. This can lead to inaccuracies both in the values of the GDR cross-sections in the high-energy region and the cross-sections with a higher reaction threshold.

In \cite{1}, attention was drawn to the discrepancy between the data on the partial cross-sections for the reactions $(\gamma,n)$, $(\gamma,2n)$, $(\gamma,3n)$ of different laboratories \cite{2,3} in the study of photo-disintegration of the $^{181}\rm{Ta}$ nucleus. It is also shown that the sums of the cross-sections $(\gamma,n)$, $(\gamma,2n)$, $(\gamma,3n)$ obtained in different laboratories agree satisfactorily. Varlamov's works \cite{4,5,6} noted systematic discrepancies between the cross-sections of partial photoneutron reactions $(\gamma,n)$, $(\gamma,2n)$, $(\gamma,3n)$ and concluded that the experimental data for many nuclei are not sufficiently reliable due to large systematic errors of the used method of separating neutrons by multiplicity. Based on experimental data and postulates of a combined model, in \cite{7,8} an experimentally-theoretical method was proposed for evaluating the values of reaction cross-sections and corrections were made between the partial photoneutron reactions on several nuclei, for example, see \cite{5,10}.

An experimental study of photonuclear reactions on stable isotopes of copper was carried out in some works \cite{11,12,13,14,15,16,17,18} and the values of the cross-sections for the reactions $(\gamma,n)$ and $(\gamma,2n)$ were obtained. The discrepancy between the data of different laboratories was observed both in the shape of the energy dependence of the cross-sections (the width of the distribution and the position of the maximum) and in the absolute value at the maximum of the cross-sections. It was shown in \cite{20} that the Fultz's data \cite{11} are systematically lower than Varlamov's ones \cite{4}, and it was pointed out that it is necessary to introduce a multiplying factor of 1.17 to the data \cite{11} on the $(\gamma,n)$ reactions for the $^{63}\rm{Cu}$ and $^{65}\rm{Cu}$ nuclei. In the review paper \cite{21}, regarding \cite{4,5}, the cross-section maxima for the reactions  ${^{63}\rm{Cu}}(\gamma,n)^{62}\rm{Cu}$, ${^{65}\rm{Cu}}(\gamma,n)^{64}\rm{Cu}$ are given: 79.79~mb and 86.38~mb, respectively, which differs significantly from results \cite{11}.
 An analysis of the experimental cross-sections on copper isotopes for the $(\gamma,2n)$ reaction from \cite{11}, carried out in \cite{5} using the experimentally-theoretical procedure \cite{7,8}, showed the need for a significant correction of these data. The authors of \cite{7,8} also believe that it is necessary to introduce corrections not only in the case of the Cu nuclei, but also for all photonuclear cross-sections obtained by the method of direct neutron detection.

Thus, we can conclude that there is a spread in the values of photonuclear cross-sections for Cu isotopes in the GDR region obtained in different
 laboratories. Considering that data on photonuclear reactions on copper isotopes sometimes are used to monitor flux of bremsstrahlung quanta, for example, ${^{63}\rm{Cu}}(\gamma,n)^{62}\rm{Cu}$ in \cite{15} and
 ${^{65}\rm{Cu}}(\gamma,n)^{64}\rm{Cu}$ in \cite{26}, it is necessary to introduce more certainty into the experimental values of the cross-sections using other methods.

In this work, using the $\gamma$-activation technique and off-line $\gamma$-ray spectrometric technique, we study photonuclear reactions ${^{65}\rm{Cu}}(\gamma,n)^{64}\rm{Cu}$, ${^{63}\rm{Cu}}(\gamma,n)^{62}\rm{Cu}$, ${^{63}\rm{Cu}}(\gamma,2n)^{61}\rm{Cu}$, ${^{63}\rm{Cu}}(\gamma,3n)^{60}\rm{Cu}$ in the bremsstrahlung end-point energy range $E_{\rm{\gamma max}}$ = 35--94 MeV. The obtained experimental results on the average cross-sections $\langle{\sigma(E_{\rm{\gamma max}})}\rangle$ of the reactions  will make it possible to supplement the data of different laboratories for the case of the reactions $(\gamma,n)$ and $(\gamma,2n)$. The cross-sections for the reaction ${^{63}\rm{Cu}}(\gamma,3n)^{60}\rm{Cu}$ have not been measured before. The analysis of the found experimental $\langle{\sigma(E_{\rm{\gamma max}})}\rangle$ is carried out using the cross-sections $\sigma(E)$ from the TALYS1.95 code \cite{27} and the data available in the literature.

\section{Flux-average cross-sections determination}
\subsection{Experimental setup and procedure}
\label{sec:1}

  Experimental study of copper photodisintegration cross-sections has been carried out through measurements of the residual $\gamma$-activity of the irradiated sample, which enabled one to obtain simultaneously the data from different channels of photonuclear reactions. This technique is well known and has been described in a variety of papers concerned with the investigation of multiparticle photonuclear reactions, e.g., on the nuclei ${^{27}\!\rm{Al}}$ \cite{28,28a}, ${^{93}\rm{Nb}}$ \cite{29,30,31}, ${^{181}\rm{Ta}}$ \cite{1,32}.

\begin{figure}[b]
   	\resizebox{0.49\textwidth}{!}{%
  \includegraphics{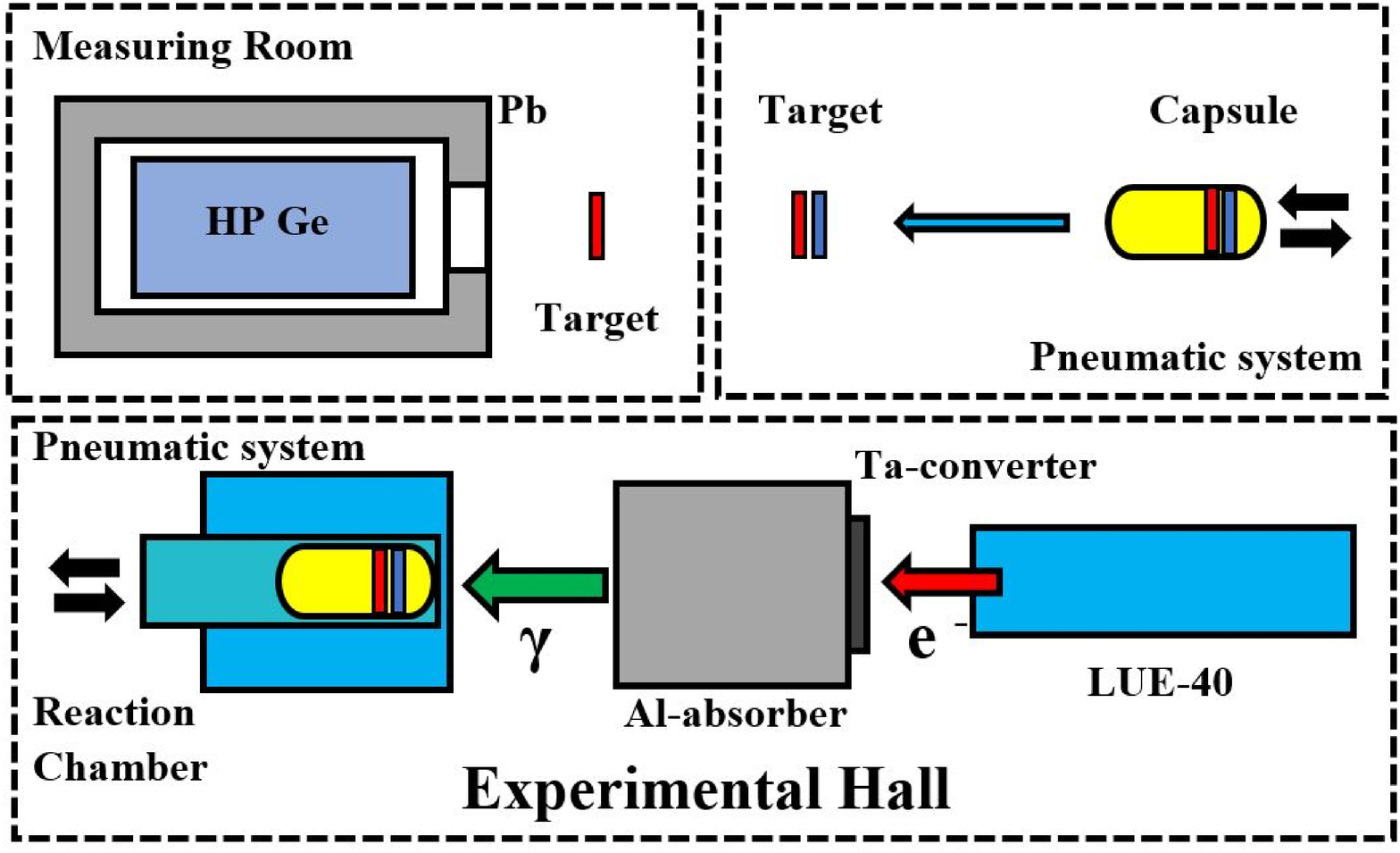}}
   	\caption{The schematic block diagram of the experimental setup. The upper part shows the measuring room, where the exposed target (red color) and the target-monitor (blue color) are extracted from the capsule and are arranged by turn before the HPGe detector for induced $\gamma$-activity measurements. The lower part shows the accelerator LUE-40, Ta-converter, Al-absorber, exposure reaction chamber.}
   	\label{fig1}
   	%\vspace{1ex}
   \end{figure}

  The schematic block diagram of the experimental setup is presented in Fig.~\ref{fig1}. The $\gamma$-ray bremsstrahlung beam was generated by means of the NSC KIPT electron linac LUE-40 RDC “Accelerator” \cite{33,34}. Electrons of the initial energy $E_e$ were incident on the target-converter made from 1.05~mm thick natural tantalum plate, measuring 20 by 20~mm. To remove electrons from the bremsstrahlung flux, a cylindrical aluminum absorber, 100~mm in diameter and 150 mm in length, was used. The targets of diameter 8~mm, placed in the aluminum capsule, were arranged behind the Al-absorber on the electron beam axis. For transporting the targets to the place of irradiation and back for induced activity registration, the pneumatic tube transfer system was used. After the irradiated targets are delivered to the measuring room, the samples are extracted from the aluminum capsule and are transferred one by one to the detector for the measurements.

The induced $\gamma$-activity of the irradiated targets was registered by the semiconductor HPGe detector Canberra GC-2018 with the resolutions of 0.8 and 1.8~keV
 (FWHM) for the $\gamma$-quanta energies $E_\gamma$ = 122 and 1332~keV, respectively. Its efficiency was 20\% relative to the NaI(Tl) detector, 3~inches in diameter and 3 inches in thickness.
   The absolute registration efficiency of the HPGe detector was calibrated with a standard set of gamma-ray radiation sources: $^{22}$Na, $^{60}$Co, $^{133}$Ba, $^{137}$Cs, $^{152}$Eu, $^{241}$Am.

The bremsstrahlung spectra of electrons were calculated by using the GEANT4.9.2 code \cite{35} with due regard for the real geometry of the experiment, where consideration was given to spatial and energy distributions of the electron beam. The program code GEANT4.9.2, PhysList {\it{G4LowEnergy}} allows one to perform calculations taking properly into account all physical processes for the case of an amorphous target. Similarly, GEANT4.9.2, PhysList {\it{QGSP-BIC-HP}} makes it possible to calculate the neutron yield due to photonuclear reactions from targets of different thicknesses and atomic charges. In addition, the bremsstrahlung gamma fluxes were monitored by the yield of the $^{100}\rm{Mo}(\gamma,n)^{99}\rm{Mo}$ reaction. For this purpose, the natural molybdenum target-witness, placed close by the target under study, was simultaneously exposed to radiation.

The $^{\rm nat}$Cu and $^{\rm nat}$Mo targets were used in the experiment. The isotopic composition of copper is a mixture of two stable isotopes: $^{63}\rm{Cu}$ (isotopic abundance 69.17\%) and $^{65}\rm{Cu}$ (isotopic abundance 30.83\%). For $^{100}\rm{Mo}$ in our calculations, we have used the percentage value of isotope abundance equal to 9.63\% (see ref.~\cite{35}). The admixture of other elements in the targets did not exceed 0.1\% by weight.

In the experiment, Cu and Mo samples were exposed to radiation at end-point bremsstrahlung energies $E_{\rm{\gamma max}}$ ranging from 35 to 94~MeV with an energy step of $\sim$5~MeV. The masses of the Cu and Mo targets were, respectively, 22~mg and 60~mg. The time of irradiation $t_{\rm irr}$ was 30~min for each energy $E_{\rm{\gamma max}}$ value; the time of residual $\gamma$-activity spectrum measurement $t_{\rm meas}$ was from 30~min to 17--60~h.

 \begin{table*}[ht]
	\caption{\label{tab1} Nuclear spectroscopic data of the radio-nuclides reactions from \cite{36}}
	\centering
%	\begin{ruledtabular}
	\begin{tabular}{ccccc}
		\hline \hline	\vspace{1ex}
		Nuclear reaction & $E_{\rm{th}}$,~MeV & $T_{1/2}$ & $E_{\gamma}$,~keV & $I_{\gamma}$, \% \\ \hline 	
		\vspace{1ex}	
		${^{65}\rm{Cu}}(\gamma,n)^{64}\rm{Cu}$  & 9.91 & 12.700 $\pm$ 0.002 h & 1345.84 & 0.473 $\pm$ 0.010 \\ \hline      	  \vspace{1ex}	
		
		${^{63}\rm{Cu}}(\gamma,n)^{62}\rm{Cu}$  & 10.86 & 9.74 $\pm$ 0.02 min & \begin{tabular}{c}1172.9\\875.68
		\end{tabular}
		& \begin{tabular}{c} 0.34\\ 0.150 $\pm$ 0.007
		\end{tabular} \\ \hline      	  \vspace{1ex}	
		
		${^{63}\rm{Cu}}(\gamma,2n)^{61}\rm{Cu}$  & 19.74 & 3.333 $\pm$ 0.005 h & \begin{tabular}{c}1185.23\\282.956\\656.008
		\end{tabular}
		& \begin{tabular}{c} 3.75 $\pm$ 0.07\\ 12.2 $\pm$ 0.3 \\ 10.77 $\pm$  0.18
		\end{tabular} \\ \hline      	  \vspace{1ex}	
		${^{63}\rm{Cu}}(\gamma,3n)^{60}\rm{Cu}$  & 31.44 & 23.7 $\pm$ 0.4 min & 1332.5 & 88  \\ \hline
		\vspace{1ex}	
		$^{100}\rm{Mo}(\gamma,n)^{99}\rm{Mo}$ & 8.29 & $65.94 \pm 0.01$~h &
		739.50 &       	$12.13 \pm 0.12$ \\	    \hline \hline
	\end{tabular}	
	\\ {The error of intensity $I_{\gamma}$ for the 1172.9 keV $\gamma$-line was determined as half-value spreads according to the databases of \cite{36} and \cite{37}. In the case of the 1332.5~keV $\gamma$-line, the $I_{\gamma}$-error is absent in \cite{36,37}, and therefore it was taken to be 0.5\%.}
%\end{ruledtabular}	
\end{table*}

The yield and bremsstrahlung flux-averaged cross-sections $\langle{\sigma(E_{\rm{\gamma max}})}\rangle$ of the $^{100}\rm{Mo}(\gamma,n)^{99}\rm{Mo}$, ${^{65}\rm{Cu}}(\gamma,n)^{64}\rm{Cu}$, ${^{63}\rm{Cu}}(\gamma,n)^{62}\rm{Cu}$, ${^{63}\rm{Cu}}(\gamma,2n)^{61}\rm{Cu}$, ${^{63}\rm{Cu}}(\gamma,3n)^{60}\rm{Cu}$ reactions were obtained. Table~\ref{tab1} lists the nuclear spectroscopic data of the radionuclide's reactions according to the data from \cite{36}: $E_{\rm{th}}$ denotes reaction thresholds; $T_{1/2}$ -- the half-life period of the nuclei-products, $E_{\gamma}$ are the energies of the $\gamma$-lines under study and their intensities $I_{\gamma}$.

Measuring the yield of photoneutron reactions on copper isotopes immediately after the end of irradiation is hampered by the high intensity of the 511 keV $\gamma$-line from positron annihilation. This can lead to random coincidences in the detection system, which can distort the result of target activity measurements. The presence of the positron line reduces the convenience of measurements, which should be attributed to the disadvantages of using the Cu nuclei as a monitor target.

In this regard, it is necessary either to significantly (up to 400 mm) increase the distance between the detector and the irradiated target or to wait several hours for the intense of short-lived residual radiation lines to weaken. We used both variants of measurements.

\begin{figure*}[]
	\resizebox{0.99\textwidth}{!}{%
		\includegraphics{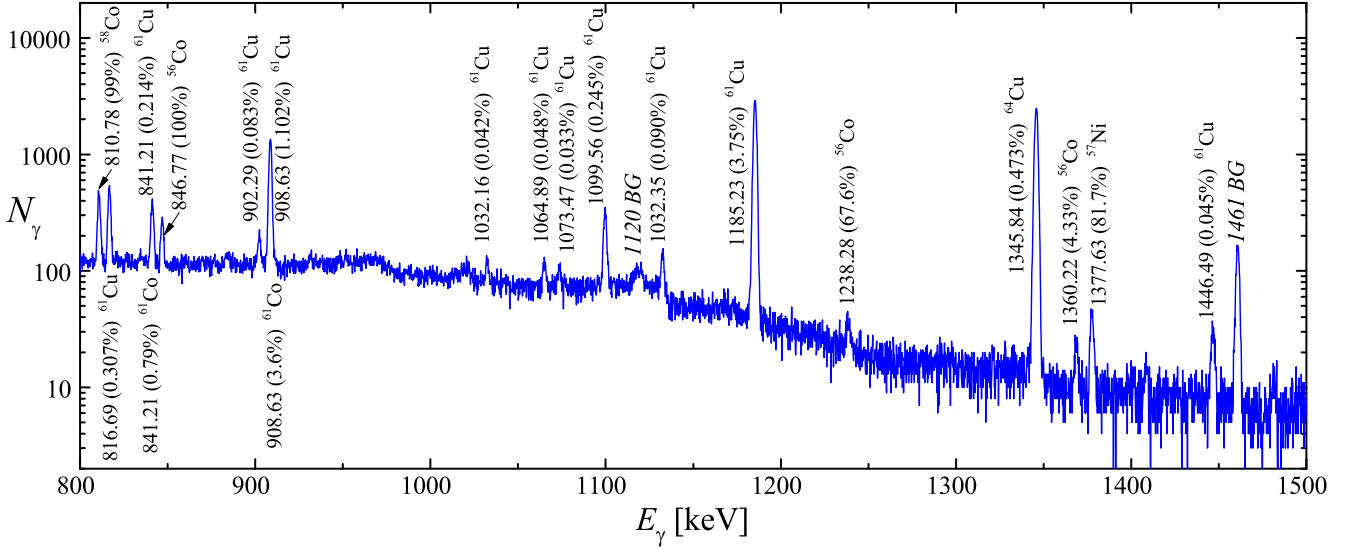}}
	\caption{Fragment of the gamma-ray spectrum from the $^{\rm nat}$Cu target measured for the following parameters: $t_{\rm meas}$ = 63218~s,  $t_{\rm cool}$ = 19015~s, $E_{\rm{\gamma max}}$ = 94~MeV, $m$ = 21.815~mg.  The spectrum fragment ranges from 800 to 1500~keV. The background $\gamma$-lines peaks  are indicated by the letters $BG$.}
	\label{fig2}
	%\vspace{1ex}
\end{figure*}

To process the spectra and estimate the number of counts of $\gamma$-quanta in the full absorption peak $\triangle A$, we used the program InterSpec v.1.0.9 \cite{38}.
Figure~\ref{fig2} shows the typical gamma-spectrum from reaction products of the copper target in the $E_{\gamma}$ range from 800 to 1500 keV.

The bremsstrahlung gamma-flux monitoring against the $^{100}\rm{Mo}(\gamma,n)^{99}\rm{Mo}$ reaction yield was performed by comparing the experimentally obtained average cross-section values with the computation data. To determine the experimental $\langle{\sigma(E_{\rm{\gamma max}})}\rangle$  values we have used the $\triangle A$ activity value for the $E_{\gamma}$ = 739.50 keV $\gamma$-line and the absolute intensity $I_{\rm \gamma}$ = 12.13\% (see Table~\ref{tab1}). The theoretical values of the average cross-section
 $\langle{\sigma(E_{\rm{\gamma max}})}\rangle_{\rm{th}}$ were calculated using the cross-sections $\sigma(E)$ from the TALYS1.95 code with the default options. The normalization (monitoring) factor $k_{\rm{monitor}}$, derived from the ratios of
   $\langle{\sigma(E_{\rm{\gamma max}})}\rangle_{\rm{th}}$ to $\langle{\sigma(E_{\rm{\gamma max}})}\rangle$, represent the deviation of the GEANT4.9.2-computed bremsstrahlung $\gamma$-flux from the real flux falling on the target.
     The determined $k_{\rm{monitor}}$ values were used for normalizing the cross-sections for other photonuclear reactions. The monitoring procedure has been detailed in \cite{29,30}.

The Ta-converter and Al-absorber, used in the experiment, generate the neutrons that can cause the reaction \\ $^{100}\rm{Mo}(n,2n)^{99}\rm{Mo}$. Calculations were made of the neutron energy spectrum and the fraction of neutrons of energies above the threshold of this reaction, similarly to \cite{39}.
 The contribution of the $^{100}\rm{Mo}(n,2n)^{99}\rm{Mo}$ reaction to the value of the induced activity of the $^{99}$Mo nucleus has been estimated and it has been shown that this contribution is negligible compared to the contribution of  $^{100}\rm{Mo}(\gamma,n)^{99}\rm{Mo}$. The contribution of the reaction $^{100}\rm{Mo}(\gamma,p)^{99}\rm{Nb}$, $^{99}\rm{Nb} \xrightarrow{\beta^-}$$^{99}\rm{Mo}$ is also negligible.

\subsection{Experimental accuracy of flux-average cross-sections $\langle{\sigma(E_{\rm{\gamma max}})}\rangle$}
\label{sec:2}

     The uncertainty in measurements of experimental values of the average cross-sections $\langle{\sigma(E_{\rm{\gamma max}})}\rangle$ was determined as a quadratic sum of statistical and systematical errors. The statistical error in the observed $\gamma$-activity is mainly due to statistics in the total absorption peak of the corresponding $\gamma$-line, which varies within 1 to 10\%. This error varies depending on the $\gamma$-line intensity and the background conditions of spectrum measurements. The observed activity $\triangle A$ of the investigated $\gamma$-line depends on the detection efficiency, the half-life period, and the absolute intensity $I_{\rm \gamma}$. The background is generally governed by the contribution of the Compton scattering of quanta.

The systematic errors are associated with the uncertainties of the: 1. irradiation time $\sim$0.5\%; 2. the electron current $\sim$0.5\%; 3. the detection efficiency of the detector 2--3\%, which is mainly associated with the error of the reference sources of $\gamma$-radiation and the choice of the approximation curve; 4. the half-life  $T_{1/2}$ of the reaction products and the absolute intensity $I_{\rm \gamma}$ of the analyzed $\gamma$-quanta, which is noted in Table~\ref{tab1}; 5. normalization of the experimental data to the  $^{100}\rm{Mo}(\gamma,n)^{99}\rm{Mo}$  monitor reaction yield up to 5\%; 6. error in calculating the contribution from competing$\gamma$-lines (described in the text).

     Thus, the statistical and systematical errors are the variables, which differ for different ${^{65}\rm{Cu}}(\gamma,n)^{64}\rm{Cu}$ and ${^{63}\rm{Cu}}(\gamma,xn)^{63-x}\rm{Cu}$ reactions. The total uncertainty of the experimental data is given in figures with experimental results.

     \subsection{Calculation formulas for flux-average cross-sections}
     \label{sec:3}

     The cross-sections $\sigma(E)$, averaged over the bremsstrahlung $\gamma$-flux $W(E,E_{\rm{\gamma max}})$ from the threshold $E_{\rm{th}}$ of the reaction under study to the end-point energy of the spectrum $E_{\rm{\gamma max}}$, were calculated with the use of the theoretical cross-section values computed with the TALYS1.95 code \cite{27}, installed on the Ubuntu20.04. The bremsstrahlung flux-averaged cross-section $\langle{\sigma(E_{\rm{\gamma max}})}\rangle_{\rm th}$ in a given energy interval was calculated by the formula:

     \begin{equation}\label{form1}
     \langle{\sigma(E_{\rm{\gamma max}})}\rangle_{\rm th} = \frac
     {\int\limits_{E_{\rm{th}}}^{E_{\rm{\gamma max}}}\sigma(E)\cdot W(E,E_{\rm{\gamma max}})dE}
     {\int\limits_{E_{\rm{th}}}^{E_{\rm{\gamma max}}}W(E,E_{\rm{\gamma max}})dE}.
     \end{equation}

     These theoretical average cross-sections were compared with the experimental values calculated by the formula:

     \begin{equation}\begin{split}
     \langle{\sigma(E_{\rm{\gamma max}})}\rangle = \\
     \frac{\lambda \triangle A  {\rm{\Phi}}^{-1}(E_{\rm{\gamma max}})}{N_x I_{\gamma} \ \varepsilon (1-\exp(-\lambda t_{\rm{irr}}))\exp(-\lambda t_{\rm{cool}})(1-\exp(-\lambda t_{\rm{meas}}))},
     \label{form2}
     \end{split}
     \end{equation}
          where $\triangle A$ is the number of counts of $\gamma$-quanta in the full absorption peak (for the $\gamma$-line of the investigated reaction), $\lambda$ is the decay constant \mbox{($\rm{ln}2/\textit{T}_{1/2}$)}; $N_x$ is the number of target atoms, $I_{\gamma}$ is the absolute intensity of the analyzed $\gamma$-quanta,
     $\varepsilon$ is the absolute detection efficiency for the analyzed photon energy,
      ${\rm{\Phi}}(E_{\rm{\gamma max}}) = {\int\limits_{E_{\rm{th}}}^{E_{\rm{\gamma max}}}W(E,E_{\rm{\gamma max}})dE}$  is the integrated bremsstrahlung flux in the energy range from the reaction threshold $E_{\rm{th}}$ up to $E_{\rm{\gamma max}}$, $t_{\rm{irr}}$, $t_{\rm{cool}}$ and $t_{\rm{meas}}$ are the irradiation time, cooling time and measurement time, respectively.
     A more detailed description of all the calculation procedures necessary for the determination of $\langle{\sigma(E_{\rm{\gamma max}})}\rangle$ can be found in \cite{29,30}.

 \section{	RESULTS AND DISCUSSIONS}
 \subsection{ Photonuclear reaction ${^{65}\rm{Cu}}(\gamma,n)^{64}\rm{Cu}$}
 \label{sec:4}

 In the ${^{65}\rm{Cu}}(\gamma,n)$ reaction the $^{64}$Cu nucleus is formed with half-life $T_{1/2}$ = 12.700$\pm$0.002~h. The $^{64}$Cu nucleus undergoes $\beta^-$-decay (39\%) with the formation of the $^{64}$Zn nucleus and
  $\varepsilon^+\beta^+$-decay (61\%) with the formation of the $^{64}$Ni nucleus. In the second
     case, a gamma-line emission with $E_{\rm{\gamma}}$ = 1345.84 keV of low-intensity
  	  	 $I_{\gamma}$ = 0.473\% is observed.

\begin{figure}[t]
  	\resizebox{0.49\textwidth}{!}{%
  \includegraphics{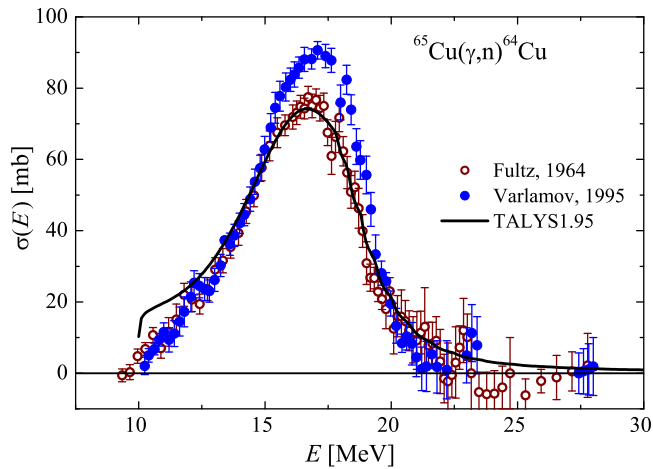}}
	\caption{The ${^{65}\rm{Cu}}(\gamma,n)^{64}\rm{Cu}$ reaction cross-section. Experimental data of Fultz \cite{11} (empty circles) and Varlamov \cite{4,21} (filled circles), curve -- calculation in TALYS1.95 code.}
	\label{fig31}
%	\vspace{1ex}
\end{figure}

An experimental study of the cross-section for the reaction ${^{65}\rm{Cu}}(\gamma,n)^{64}\rm{Cu}$ was carried out in several works, for example, \cite{11} and \cite{4,21}. The results of these works have significant differences in the value of the GDR maxima: 75$\pm$7~mb \cite{11}, 86.38~mb \cite{21} (see Fig.~\ref{fig31}). These data were matched in the region of the GDR maximum in \cite{4} by introducing a coefficient of 1.17 into the data of \cite{11}.

The calculations of the cross-section $\sigma(E)$ for this reaction were carried out in the TALYS1.95 code with default parameters. As can be seen from Fig.~\ref{fig31}, the calculated cross-section at the maximum is very close to the experimental values of $\sigma(E)$  from \cite{11}, but significantly lower than Varlamov's data \cite{4,21}. The leading edge of the calculated distribution does not coincide with the data of the two experiments.

\begin{figure}[t]
  	\resizebox{0.49\textwidth}{!}{%
  \includegraphics{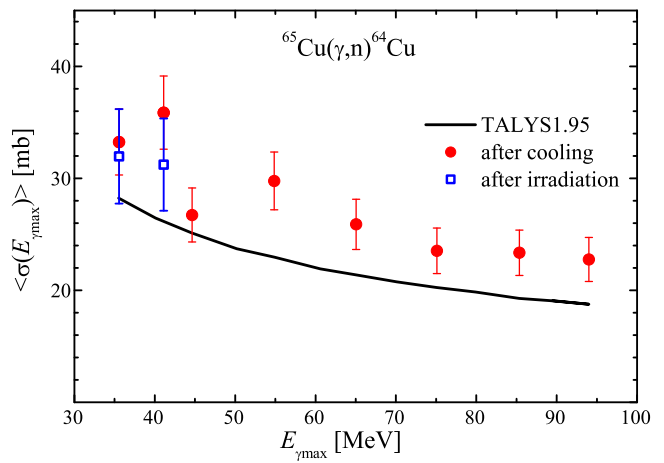}}
	\caption{Flux-average cross-section $\langle{\sigma(E_{\rm{\gamma max}})}\rangle$ of the reaction ${^{65}\rm{Cu}}(\gamma,n)^{64}\rm{Cu}$. The curve is calculated using the TALYS1.95 code. Experimental result: squares -- measurements immediately after irradiation, circles -- measurements after cooling.}
	\label{fig32}
%	\vspace{1ex}
\end{figure}

The average cross-sections $\langle{\sigma(E_{\rm{\gamma max}})}\rangle$ calculated using the cross-section from TALYS1.95 and our experimental data for the reaction ${^{65}\rm{Cu}}(\gamma,n)^{64}\rm{Cu}$ are shown in Fig.~\ref{fig32}. It can be seen that all experimental points are located noticeably higher than the calculated curve ($\sim$15--20\%), which is evidence in favor of the data of \cite{4,21}.

	\subsection{ Photonuclear reaction  ${^{65}\rm{Cu}}(\gamma,2n)^{63}\rm{Cu}$}
\label{sec:5}

\begin{figure}[h]
  	\resizebox{0.49\textwidth}{!}{%
  \includegraphics{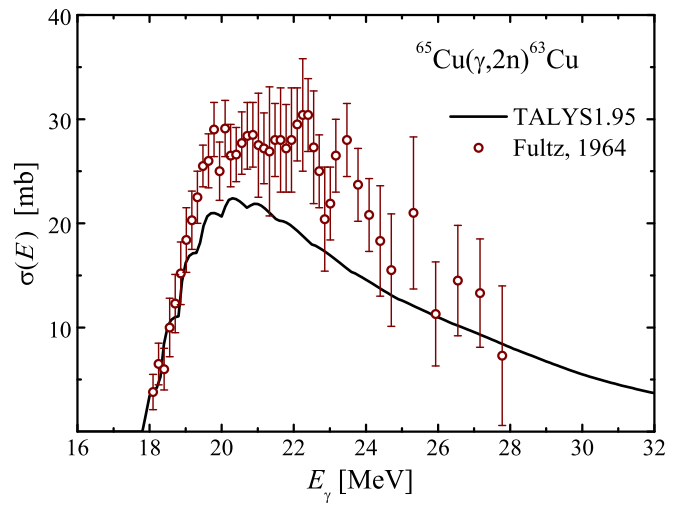}}
	\caption{The ${^{65}\rm{Cu}}(\gamma,2n)^{63}\rm{Cu}$ reaction cross-section. Curve -- calculation in the TALYS1.95 code, points -- experimental data from \cite{11}.}
	\label{fig33}
	\vspace{2ex}
\end{figure}

  As a result of the ${^{65}\rm{Cu}}(\gamma,2n)$ reaction, the stable isotope $^{63}$Cu is formed. The induced activity method does not allow us to measure the yield of this reaction. For completeness of the analysis, Fig.~\ref{fig33} shows the data from \cite{11} and calculation from TALYS1.95 with default parameters. It can be seen that there is a noticeable difference between the experimental and calculated cross-sections. The experiment exceeds the theoretical cross-sections from the TALYS1.95 code by approximately 1.3 times at energies of 19--24~MeV.

  In \cite{5}, using the experimental-theoretical approach \cite{7,8}, the reliability of data on the cross-sections for the reaction ${^{65}\rm{Cu}}(\gamma,2n)^{63}\rm{Cu}$ from \cite{11}  was verified. It is shown that the estimated (recalculated) values of the cross-sections are significantly lower than the experimental ones, and the difference was up to 30\% in the energy range $E_{\rm \gamma}$ = 19--25~MeV. This value is very close to the discrepancy between the TALYS1.95 cross-section and the result \cite{11}.

  Such a significant change (correction) in the cross-section for the ${^{65}\rm{Cu}}(\gamma,2n)^{63}\rm{Cu}$ reaction leads to corrections in the GDR cross-section for energies above 18~MeV.

\subsection{ Photonuclear reaction  ${^{63}\rm{Cu}}(\gamma,n)^{62}\rm{Cu}$}
\label{sec:6}

In the ${^{63}\rm{Cu}}(\gamma,n)$ reaction the $^{62}\rm{Cu}$ nucleus is formed with half-life $T_{1/2}$ = 9.74~m. The $^{62}\rm{Cu}$ nucleus undergoes $\varepsilon^+\beta^+$-decay (the $^{62}\rm{Ni}$ nucleus is formed) with the emission of several low-intensity gamma-lines. The line of $\gamma$-radiation with $E_{\rm \gamma}$ = 1172.9~keV and intensity $I_{\rm \gamma}$ = 0.34\% was used in the work.

An experimental study of this reaction was carried out in several works \cite{11,12,13,14,15} and \cite{4,21} in the energy range 10--28 MeV. These data differ in the height of the GDR maximum: the data of \cite{11} are lower than the results of other laboratories, as in the case of the ${^{65}\rm{Cu}}(\gamma,n)^{64}\rm{Cu}$ reaction. The disagreement in the GDR maxima between \cite{11} and \cite{4,21} was 14\%, respectively, 70$\pm$7~mb and 79.8~mb.

\begin{figure}[]
  	\resizebox{0.49\textwidth}{!}{%
  \includegraphics{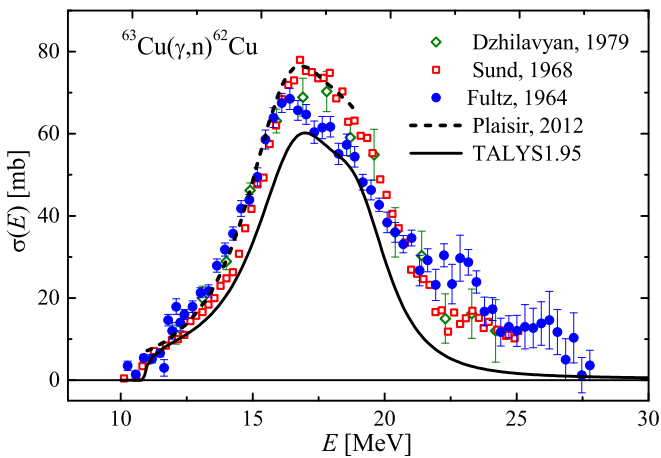}}
	\caption{The ${^{63}\rm{Cu}}(\gamma,n)^{62}\rm{Cu}$ reaction cross-section. Calculated cross-sections $\sigma(E)$ from the Talys1.95 code (solid curve) and experimental data: empty diamonds -- \cite{13},  empty squares -- \cite{12}, filled circles -- \cite{11}, dashed curve -- \cite{15}.}
	\label{fig34}
%	\vspace{2ex}
\end{figure}

There is calculation of cross-section in the TALYS1.95 code with default parameters shown  in Fig.~\ref{fig34}. In the region of the maximum, there is good agreement between the TALYS1.95 calculation and the data of \cite{11}. All other experimental results \cite{12,13,14,15}, including \cite{21}, are located above the theoretical estimate. Note that the calculated distribution widths $\sigma(E)$ are noticeably smaller than any experimental data set.

\begin{figure}[]
  	\resizebox{0.49\textwidth}{!}{%
  \includegraphics{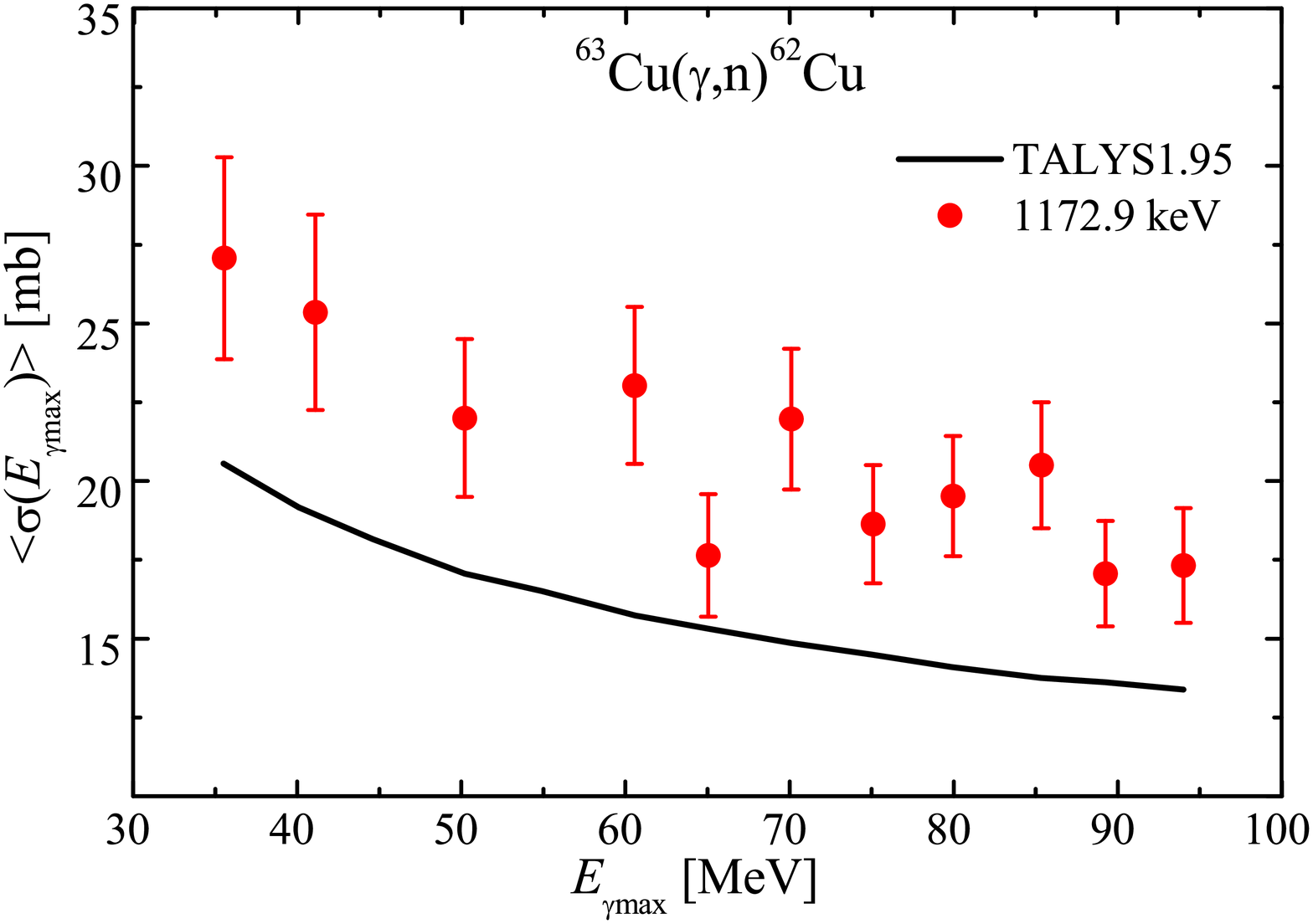}}
	\caption{Flux-average cross-sections $\langle{\sigma(E_{\rm{\gamma max}})}\rangle$ for the reaction ${^{63}\rm{Cu}}(\gamma,n)^{62}\rm{Cu}$. Curve -- calculation using Talys1.95, filled circles -- our experimental data.}
	\label{fig35}
	%\vspace{2ex}
\end{figure}

There are the average cross-sections $\langle{\sigma(E_{\rm{\gamma max}})}\rangle$ of the ${^{63}\rm{Cu}}(\gamma,n)^{62}\rm{Cu}$ reaction calculated in the TALYS1.95 code with default parameters and the obtained experimental data in the energy range $E_{\rm{\gamma max}}$ = 35--94 MeV shown in Fig.~\ref{fig35}. There is an excess of experimental $\langle{\sigma(E_{\rm{\gamma max}})}\rangle$ over the calculation for the entire energy range and is to be 15--20\%. This result agrees with that observed in Fig.~\ref{fig34} discrepancy between data \cite{12,13,14,15} and cross-sections from TALYS1.95.

Because natural copper targets were used in the study, the effect of the ${^{65}\rm{Cu}}(\gamma,3n)^{62}\rm{Cu}$ reaction on the yield of the ${^{63}\rm{Cu}}(\gamma,n)^{62}\rm{Cu}$ reaction was estimated. For this purpose, the values of the cross-section from the TALYS1.95 code were used to calculate the yield. It was found that the contribution of ${^{65}\rm{Cu}}(\gamma,3n)^{62}\rm{Cu}$  increases with increasing energy $E_{\rm{\gamma max}}$ from 0.16\% at 35 MeV to 1.32\% at 94 MeV.

In the $^{\rm nat}$Cu target, under the impact of high-energy bremsstrahlung, isotopes of cobalt and copper are also formed, which have $\gamma$-radiation with $E_{\rm{\gamma}}$ value close to that used in the study. Thus, $^{62}$Co ($E_{\rm{\gamma}}$ = 1172.9 keV, $I_{\rm{\gamma}}$ = 84\%, $T_{1/2}$ = 1.5 min), $^{62m}$Co ($E_{\rm{\gamma}}$ = 1172.9 keV, $I_{\rm{\gamma}}$ = 98\%, $T_{1/2}$ = 13.91 min), $^{60}$Co ($E_{\rm{\gamma}}$ = 1173.24 keV, $I_{\rm{\gamma}}$ = 99.974\%, $T_{1/2}$ = 5.27 years), $^{60}$Cu ($E_{\rm{\gamma}}$ = 1173.24 keV, $I_{\rm{\gamma}}$ = 0.26\%, $T_{1/2}$ = 23.7 min).

The contributions of such $\gamma$-radiation were estimated using the cross-sections from the TALYS1.95 code, taking into account the half-lives, the intensities of the competing lines, and the experimental conditions (irradiation, cooling, and measurement times). The calculated total activity of $^{62}$Co and $^{62m}$Co nuclei ranged from 0 to 1.72\% in the range of studied $E_{\rm{\gamma max}}$. The contribution of radiation from radionuclides $^{60}$Cu and $^{60}$Co is insignificant.

The total amount of corrections ranged from 0.24--3.28\% and was taken into account in the data.

\subsection{ Photonuclear reaction ${^{63}\rm{Cu}}(\gamma,2n)^{61}\rm{Cu}$}
\label{sec:7}

In the reaction ${^{63}\rm{Cu}}(\gamma,2n)$ the $^{61}\rm{Cu}$ nucleus is formed with half-life  $T_{1/2}$ = 3.333 h. The  $^{61}\rm{Cu}$  nucleus undergoes $\varepsilon^+\beta^+$-decay with the formation of the  $^{61}\rm{Ni}$ nucleus. Three lines with $E_{\rm \gamma}$ = 282.956, 656.008, and 1185.234~keV and intensity $I_{\rm \gamma}$  of 12.2, 10.77, and 3.75\%, respectively, are used for the study.

\begin{figure}[t]
  	\resizebox{0.49\textwidth}{!}{%
  \includegraphics{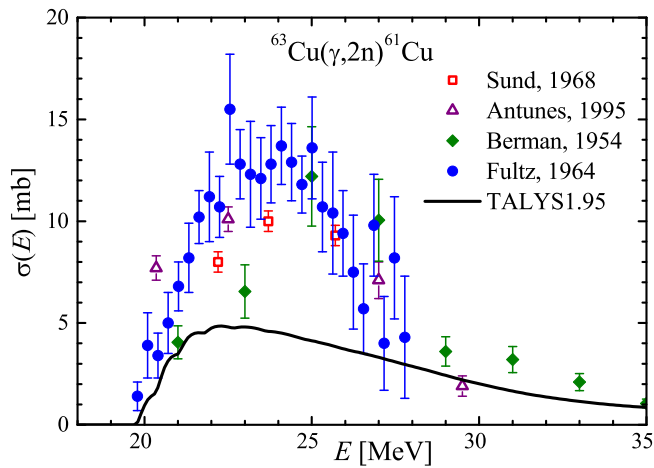}}
	\caption{The ${^{63}\rm{Cu}}(\gamma,2n)^{61}\rm{Cu}$ reaction cross-section. Calculated cross-sections $\sigma(E)$ from the Talys1.95 code (curve) and experimental data: empty squares -- \cite{12}, empty triangles -- \cite{14}, filled diamonds -- \cite{16}, filled circles -- \cite{11}.}
	\label{fig36}
%	\vspace{2ex}
\end{figure}

The calculations of the cross-sections for the ${^{63}\rm{Cu}}(\gamma,2n)$ reaction, performed in the Talys1.95 code with default parameters, are shown in Fig.~\ref{fig36}. All experimental results \cite{11,14,16}  are significantly higher than the theoretical estimates: the difference between the data \cite{11} and the calculated cross-sections reaches 3 times at the cross-section maximum. The energy dependence of the calculated curve is wider than the experiment shows.

\begin{figure}[t]
  	\resizebox{0.49\textwidth}{!}{%
  \includegraphics{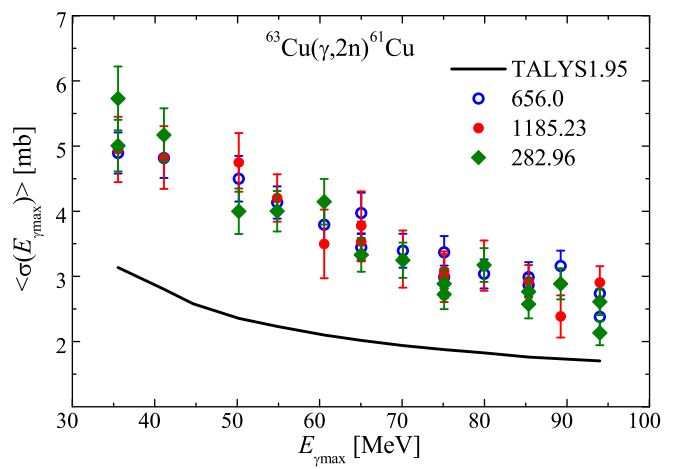}}
	\caption{Flux-average cross-sections $\langle{\sigma(E_{\rm{\gamma max}})}\rangle$ of the ${^{63}\rm{Cu}}(\gamma,2n)^{61}\rm{Cu}$ reaction. The calculation using Talys1.95 code is denoted by solid curve, our experimental data: filled diamonds -- $E_{\rm{\gamma}}$ = 282.956~keV, empty circles -- 656.008~keV, filled circles -- 1185.234~keV.}
	\label{fig37}
	%\vspace{2ex}
\end{figure}

The experimental average cross-sections $\langle{\sigma(E_{\rm{\gamma max}})}\rangle$ for the reaction${^{63}\rm{Cu}}(\gamma,2n)^{61}\rm{Cu}$ in the energy range $E_{\rm{\gamma max}}$ = 35--94 MeV are shown in Fig.~\ref{fig37}. The data found for the three gamma lines agree within the experimental error. To take into account the contribution of the ${^{65}\rm{Cu}}(\gamma,4n)^{61}\rm{Cu}$ reaction to the yield of the ${^{63}\rm{Cu}}(\gamma,2n)^{61}\rm{Cu}$ reaction, calculations were carried out similarly to the ${^{63}\rm{Cu}}(\gamma,n)^{62}\rm{Cu}$  reaction. The values of the yield for ${^{65}\rm{Cu}}(\gamma,4n)^{61}\rm{Cu}$ was calculated and it was shown that this contribution increases with energy: for 50~MeV it does not exceed 0.7\%, and for 94 MeV it reaches 3.2\%. The cross-section for the ${^{63}\rm{Cu}}(\gamma,2n)^{61}\rm{Cu}$  reaction was corrected (reduced) by the corresponding values at each energy $E_{\rm{\gamma max}}$.

The comparison the experimental cross-sections \\ $\langle{\sigma(E_{\rm{\gamma max}})}\rangle$  and the calculation with the Talys1.95 code is shown in Fig.~\ref{fig37}. As can be seen from the figure, the experimental $\langle{\sigma(E_{\rm{\gamma max}})}\rangle$  exceeds the calculated values over the entire energy range $E_{\rm{\gamma max}}$ and it averaged $\sim$100\%. 

In \cite{5}, the validity of the data for the ${^{63}\rm{Cu}}(\gamma,2n)^{61}\rm{Cu}$ reaction from \cite{11} was verified. As in the case of the \\ ${^{65}\rm{Cu}}(\gamma,2n)^{63}\rm{Cu}$  reaction, it was shown that the estimated (recalculated) data are significantly lower than the experimental data \cite{11} -- approximately by a factor of 1.5 at the energy $E$ = 23 MeV. However, such a significant change in the data from \cite{11} does not lead to their agreement with the calculation in Talys1.95, the difference between the calculation and the recalculated data was about 2 times. This value is very close to the discrepancy between the calculation and the data found in this work.

\subsection{ Photonuclear reaction ${^{63}\rm{Cu}}(\gamma,3n)^{60}\rm{Cu}$}
\label{sec:8}

In the reaction ${^{63}\rm{Cu}}(\gamma,3n)$ the $^{60}\rm{Cu}$ nucleus is formed, the half-life of which is $T_{1/2}$ = 23.7 min. The $^{60}\rm{Cu}$  nucleus undergoes $\varepsilon^+\beta^+$-decay with the formation of the $^{60}\rm{Ni}$  nucleus. To study this reaction, a $\gamma$-line with $E_{\rm \gamma}$ = 1332.5 keV and intensity of $I_{\rm \gamma}$ = 88\% is used.

\begin{figure}[]
  	\resizebox{0.49\textwidth}{!}{%
  \includegraphics{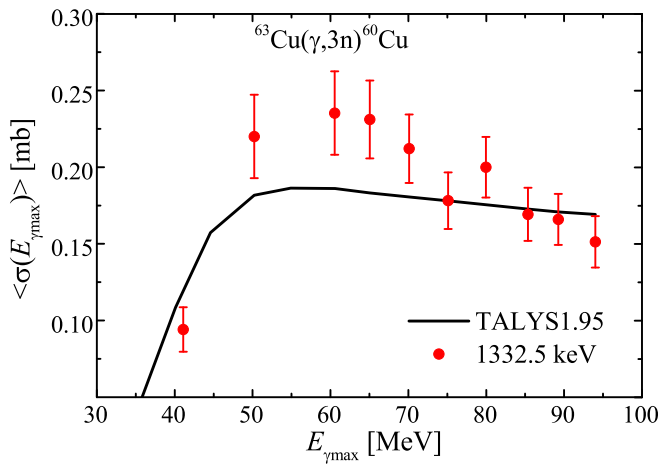}}
	\caption{The calculated average cross-sections $\langle{\sigma(E_{\rm{\gamma max}})}\rangle$ for reaction ${^{63}\rm{Cu}}(\gamma,3n)^{60}\rm{Cu}$  from the Talys1.95 code (curve) and the experimental data measured in this work (filled circles).}
	\label{fig38}
%	\vspace{2ex}
\end{figure}

Fig.~\ref{fig38} shows the calculated average cross-sections \\$\langle{\sigma(E_{\rm{\gamma max}})}\rangle$  for reaction ${^{63}\rm{Cu}}(\gamma,3n)^{60}\rm{Cu}$  from the Talys1.95 code and the experimental data measured in this work. As can be seen, at $E_{\rm{\gamma max}}$ = 50--70 MeV, the experimental values of $\langle{\sigma(E_{\rm{\gamma max}})}\rangle$  also show an excess over the calculated ones.

The values of the yield for ${^{65}\rm{Cu}}(\gamma,5n)^{60}\rm{Cu}$  was calculated and it was shown that this contribution increases with energy: for 60 MeV it does not exceed 0.7\%, and for 94 MeV it reaches 5.1\%. As in the case of the  ${^{63}\rm{Cu}}(\gamma,n)^{62}\rm{Cu}$  reaction, it is necessary to take into account the contribution of $\gamma$-radiation from competing lines from ${^{60}\rm{Co}}$ ($E_{\rm \gamma}$ = 1332.5 keV, $I_{\rm \gamma}$  = 99.986\%, $T_{1/2}$ = 5.27 years) and ${^{60m}\rm{Co}}$ ($E_{\rm \gamma}$ = 1332.5 keV, $I_{\rm \gamma}$  = 0.24\%, $T_{1/2}$ = 10.47 min). The calculated total activity of ${^{60}\rm{Co}}$  and ${^{60m}\rm{Co}}$ nuclei ranged from 0.1 to 0.15\% in the range of studied $E_{\rm{\gamma max}}$.

The value of total corrections increases with energy and amounts to 0.1--5.3\%. The cross-section for the \\${^{63}\rm{Cu}}(\gamma,3n)^{60}\rm{Cu}$ reaction was corrected (reduced) by the corresponding values at each energy $E_{\rm{\gamma max}}$.

\section{Conclusions}
\label{Concl}

The total bremsstrahlung flux-averaged cross-sections  \\$\langle{\sigma(E_{\rm{\gamma max}})}\rangle$  for the photonuclear reactions ${^{65}\rm{Cu}}(\gamma,n)^{64}\rm{Cu}$,  ${^{63}\rm{Cu}}(\gamma,n)^{62}\rm{Cu}$, ${^{63}\rm{Cu}}(\gamma,2n)^{61}\rm{Cu}$  and ${^{63}\rm{Cu}}(\gamma,3n)^{60}\rm{Cu}$  have been measured in the range of end-point energies $E_{\rm{\gamma max}}$ = 35--94 MeV. The experiments were performed with the beam from the NSC KIPT electron linear accelerator LUE-40 with the use of the activation and off-line $\gamma$-ray spectrometric technique. The calculation of flux-average cross-sections  $\langle{\sigma(E_{\rm{\gamma max}})}\rangle_{\rm th}$  and yields were carried out using the cross-section values computed with the TALYS1.95 code with the default options.

The experimental bremsstrahlung flux-averaged cross-sections  $\langle{\sigma(E_{\rm{\gamma max}})}\rangle$  for the photonuclear reactions \\${^{65}\rm{Cu}}(\gamma,n)^{64}\rm{Cu}$,  ${^{63}\rm{Cu}}(\gamma,n)^{62}\rm{Cu}$, ${^{63}\rm{Cu}}(\gamma,2n)^{61}\rm{Cu}$ systematically higher than theoretical TALYS1.95 estimates. The obtained experimental results on  $\langle{\sigma(E_{\rm{\gamma max}})}\rangle$  supplement the data available in the literature for the case of the reactions $(\gamma,n)$ and $(\gamma,2n)$ and can be used to analyze the discrepancies in the results of different laboratories when analyzing possible corrections for reaction cross-sections obtained by direct neutron detection.

It is shown that the obtained experimental data on the reactions ${^{65}\rm{Cu}}(\gamma,n)^{64}\rm{Cu}$,  ${^{63}\rm{Cu}}(\gamma,n)^{62}\rm{Cu}$ agree satisfactorily with the results of  \cite{21}, and in the case of \\ ${^{63}\rm{Cu}}(\gamma,2n)^{61}\rm{Cu}$, with the data recalculated in  \cite{5}  from work  \cite{11}.

The data for the  $\langle{\sigma(E_{\rm{\gamma max}})}\rangle$  reaction  ${^{63}\rm{Cu}}(\gamma,3n)^{60}\rm{Cu}$  were measured for the first time.

\section*{Acknowlegment}
The authors would like to thank the staff of the linear electron accelerator LUE-40 NSC KIPT, Kharkiv, Ukraine, for their cooperation in the realization of the experiment.

\section*{Declaration of competing interest}
The authors declare that they have no known competing financial interests or personal relationships that could have appeared to influence the work reported in this paper.

\end{document}